 \documentclass[preprint2]{aastex}

\shorttitle{Gamma-ray observation of M\,87 with VERITAS}
\shortauthors{Acciari et al.}

\begin{document}

\title{Observation of gamma-ray emission from the galaxy M\,87 above 250\,GeV with VERITAS}

\author{
 V.A. Acciari\altaffilmark{1},
 M. Beilicke\altaffilmark{2},
 G. Blaylock\altaffilmark{3},
 S.M. Bradbury\altaffilmark{4},
 J.H. Buckley\altaffilmark{2},
 V. Bugaev\altaffilmark{2},
 Y. Butt\altaffilmark{5},
 O. Celik\altaffilmark{6},
 A. Cesarini\altaffilmark{7},
 L. Ciupik\altaffilmark{8},
 P. Cogan\altaffilmark{9},
 P. Colin\altaffilmark{*}\altaffilmark{10},
 W. Cui\altaffilmark{11},
 M.K. Daniel\altaffilmark{4},
 C. Duke\altaffilmark{12},
 T. Ergin\altaffilmark{3},
 A.D. Falcone\altaffilmark{13},
 S.J. Fegan\altaffilmark{6},
 J.P. Finley\altaffilmark{11},
 G. Finnegan\altaffilmark{10},
 P. Fortin\altaffilmark{14},
 L.F. Fortson\altaffilmark{8},
 K. Gibbs\altaffilmark{5},
 G.H. Gillanders\altaffilmark{7},
 J. Grube\altaffilmark{4},
 R. Guenette\altaffilmark{9},
 G. Gyuk\altaffilmark{8},
 D. Hanna\altaffilmark{9},
 E. Hays\altaffilmark{15}\,\altaffilmark{16}\,\altaffilmark{17},
 J. Holder\altaffilmark{18},
 D. Horan\altaffilmark{16},
 S.B. Hughes\altaffilmark{2},
 M.C. Hui\altaffilmark{1},
 T.B. Humensky\altaffilmark{15},
 A. Imran\altaffilmark{19},
 P. Kaaret\altaffilmark{20},
 M. Kertzman\altaffilmark{21},
 D.B. Kieda\altaffilmark{10},
 J. Kildea\altaffilmark{5},
 A. Konopelko\altaffilmark{11},
 H. Krawczynski\altaffilmark{2},
 F. Krennrich\altaffilmark{19},
 M.J. Lang\altaffilmark{7},
 S. LeBohec\altaffilmark{10},
 K. Lee\altaffilmark{2},
 G. Maier\altaffilmark{9},
 A. McCann\altaffilmark{9},
 M. McCutcheon\altaffilmark{9},
 J. Millis\altaffilmark{11},
 P. Moriarty\altaffilmark{1},
 R. Mukherjee\altaffilmark{14},
 T. Nagai\altaffilmark{19},
 R.A. Ong\altaffilmark{6},
 D. Pandel\altaffilmark{20},
 J.S. Perkins\altaffilmark{5},
 M. Pohl\altaffilmark{19},
 J. Quinn\altaffilmark{22},
 K. Ragan\altaffilmark{9},
 P.T. Reynolds\altaffilmark{23},
 H.J. Rose\altaffilmark{4},
 M. Schroedter\altaffilmark{19},
 G.H. Sembroski\altaffilmark{11},
 A.W. Smith\altaffilmark{4},
 D. Steele\altaffilmark{8},
 S.P. Swordy\altaffilmark{15},
 A. Syson\altaffilmark{4},
 J.A. Toner\altaffilmark{7},
 L. Valcarcel\altaffilmark{9},
 V.V. Vassiliev\altaffilmark{6},
 S.P. Wakely\altaffilmark{15},
 J.E. Ward\altaffilmark{22},
 T.C. Weekes\altaffilmark{5},
 A. Weinstein\altaffilmark{6},
 R.J. White\altaffilmark{4},
 D.A. Williams\altaffilmark{24},
 S.A. Wissel\altaffilmark{15},
 M.D. Wood\altaffilmark{6},
 B. Zitzer\altaffilmark{11}.
}

\altaffiltext{*}{Corresponding author: colin@physics.utah.edu}
\altaffiltext{1}{Department of Life and Physical Sciences, Galway-Mayo Institute of Technology, Galway, Ireland}
\altaffiltext{2}{Department of Physics, Washington University, St. Louis, MO 63130, USA}
\altaffiltext{3}{Department of Physics, University of Massachusetts, Amherst, MA 01003, USA}
\altaffiltext{4}{School of Physics and Astronomy, University of Leeds, Leeds, LS2 9JT, UK}
\altaffiltext{5}{Fred Lawrence Whipple Observatory, Harvard-Smithsonian Center for Astrophysics, Amado, AZ 85645, USA}
\altaffiltext{6}{Department of Physics and Astronomy, University of California, Los Angeles, CA 90095, USA}
\altaffiltext{7}{Physics Department, National University of Ireland, Galway, Ireland}
\altaffiltext{8}{Astronomy Department, Adler Planetarium and Astronomy Museum, Chicago, IL 60605, USA}
\altaffiltext{9}{Physics Department, McGill University, Montreal, QC H3A 2T8, Canada}
\altaffiltext{10}{Department of Physics, University of Utah, Salt Lake City, UT 84112, USA}
\altaffiltext{11}{Department of Physics, Purdue University, West Lafayette, IN 47907, USA}
\altaffiltext{12}{Department of Physics, Grinnell College, Grinnell, IA 50112-1690, USA}
\altaffiltext{13}{Department of Astronomy and Astrophysics, Pennsylvania State University, PA 16802, USA}
\altaffiltext{14}{Department of Physics and Astronomy, Barnard College, Columbia University, NY 10027, USA}
\altaffiltext{15}{Enrico Fermi Institute, University of Chicago, Chicago, IL 60637, USA}
\altaffiltext{16}{Argonne National Laboratory, Argonne, IL 60439, USA}
\altaffiltext{17}{now at N.A.S.A./Goddard Space-Flight Center, Greenbelt, MD 20771, USA}
\altaffiltext{18}{Department of Physics and Astronomy and the Bartol Research Institute, University of Delaware, Newark, DE 19716, USA}
\altaffiltext{19}{Department of Physics and Astronomy, Iowa State University, Ames, IA 50011, USA}
\altaffiltext{20}{Department of Physics and Astronomy, University of Iowa, Iowa City, IA 52242, USA}
\altaffiltext{21}{Department of Physics and Astronomy, DePauw University, Greencastle, IN 46135-0037, USA}
\altaffiltext{22}{School of Physics, University College Dublin, Belfield, Dublin 4, Ireland}
\altaffiltext{23}{Department of Applied Physics and Instrumentation, Cork Institute of Technology, Bishopstown, Cork, Ireland}
\altaffiltext{24}{Santa Cruz Institute for Particle Physics and Department of Physics, University of California, Santa Cruz, CA 95064, USA}

\begin{abstract}
The multiwavelength observation of the nearby radio galaxy M\,87 provides a unique opportunity to study in 
detail processes occurring in Active Galactic Nuclei from radio waves to TeV $\gamma$-rays.
Here we report the detection of $\gamma$-ray emission above 250\,GeV from M\,87 in spring 2007
with the VERITAS atmospheric Cherenkov telescope array and discuss its correlation with the X-ray emission.
The $\gamma$-ray emission is measured to be point-like with an
intrinsic source radius less than 4.5 arcmin.
The differential energy spectrum is fitted well by a power-law function: 
d$\Phi/$dE\,=$(7.4\pm1.3_{stat}\pm1.5_{sys})\times$ (E/TeV)$^{(-2.31\pm0.17_{stat}\pm0.2_{sys})} 10^{-9}$m$^{-2}$s$^{-1}$TeV$^{-1}$.
We show strong evidence for a year-scale correlation between the $\gamma$-ray flux reported by
TeV experiments and the X-ray emission measured by the ASM/RXTE observatory,
and discuss the possible short-time-scale variability.
These results imply that the $\gamma$-ray emission from M\,87 is more likely associated with
the core of the galaxy than with other bright X-ray features in the jet.
\end{abstract}


\section{Introduction}
The giant elliptical galaxy M\,87 is a nearby ($\sim$16~Mpc) powerful FR I
\citep{fan74} radio galaxy (Virgo A) which lies near the center of the Virgo
cluster. Its core is an Active Galactic Nucleus (AGN) powered by a
supermassive black hole of $(3.2\pm0.9)\times 10^9~\mathrm{M}_\odot$
\citep{mac97}, emitting the first-observed plasma jet~\citep{cur18}.
M\,87 has been observed over a broad range of energies from radio
waves to TeV $\gamma$-rays. Its jet is resolved in radio, optical and
X-ray regimes and shows similar morphologies at all these wavelengths.
M\,87 is the first non-blazar AGN observed to emit TeV $\gamma$-rays and
it provides valuable insight into the acceleration of high-energy particles
in astrophysical jets.

The first detection of M\,87 in the TeV regime was reported by the
HEGRA collaboration~\citep{aha03} with a statistical significance of
4.1 standard deviations derived from observations made during 1998-99.
This detection was confirmed by the HESS collaboration~\citep{aha06A}, which
also reported a year-scale flux variability and strong indications
of rapid variability (2-day scale) during a high state of $\gamma$-ray activity in 2005.
The imaging atmospheric Cherenkov technique provides insufficient
angular resolution (few arcminutes) to resolve the M\,87 emission region,
but the day-scale variability claimed by HESS suggests a very small region,
most likely close to the core. The confirmation of such short-time-scale variability
is of paramount importance because of its implications regarding the source dimensions.

The rapid variability of M\,87 $\gamma$-ray emission would also constitute
an important connection with the most common extragalactic sources detected in the TeV regime.
These are all generally considered to be BL Lacertae (BL Lac) objects with the jet pointing
towards the observer (blazar).
The TeV emission of M\,87 is now generally interpreted in the frame of the unified scheme
of AGN~\citep{urr95}. According to this scheme, FR I radio galaxies are of the same nature
as BL Lac objects, but with their jet not pointing along the line of sight.
Typical models assume a M\,87 jet misalignment around 30$^\circ$-35$^\circ$~\citep{bic96}.
However, superluminal motion has been observed in the compact first knot, HST-1,
in the optical \citep{bir99} and radio domains \citep{che07}. To explain this requires a jet orientation
closer to the line of sight (within 19$^\circ$) at least at this location in the jet.
The HST-1 emission also shows \lq\lq blazar-like\rq\rq~behavior with
very strong month-scale variability in the radio, optical and X-ray wavelengths \citep{har07A},
much stronger than the variability observed in the nucleus or in the other knots.

The non-thermal emission of the jet is understood as synchrotron
radiation from high-energy electrons, and the TeV $\gamma$-rays
could result from inverse-Compton scattering by the same electron population.
A time correlation between the X-ray synchrotron radiation 
and the TeV emission is then expected.
Models explain the TeV emission as originating
from electrons either in the inner jet close
to the core and the jet emission base (TeV blazar-type model) (\citet{geo05}, \citet{ghi05},
\citet{len07}, \citet{tav08}), or in the large-scale jet (\citet{sta03}, \citet{hon07}).
On intermediate scale, it was recently proposed the TeV emission comes from the peculiar
knot HST-1 \citep{sta06}.
Since the year-scale variability is established, the core region and the knot HST-1
are more likely the source of the TeV emission than the large-scale jet.

The origin of the TeV emission could also not be in the jet but in the vicinity of
the central supermassive black hole of M\,87. The rotating magnetosphere of the black hole
could accelerate particles (pulsar-type model) producing electromagnetic cascades
and then TeV $\gamma$-rays by inverse-Compton scattering (\citet{ner07}, \citet{rie08}).
This model could explain a day-scale variability of the TeV emission.

An alternative model involving protons has also been proposed \citep{rei04}.
In this model, the TeV $\gamma$-ray emission is dominated by neutral pion production,
and by proton and muon synchrotron radiation in the highly magnetized environment
of the jet formation region (Synchrotron-Proton-Blazar model).
In this context, M\,87 would be an efficient cosmic rays accelerator.
It was even suggested that M\,87 is the source of most ultra-high-energy
cosmic rays detected on Earth \citep{bie01}.
However the SPB model predicts a quite soft spectrum in the TeV range,
softer than what was recently measured by HESS and VERITAS.

Another model, involving dark matter annihilation, has also predicted 
TeV $\gamma$-ray emission from M\,87 \citep{bal00}. But the variability
of the flux almost excludes this scenario as the main contributer to
the TeV emission. Dark matter is expected to be distributed over
much larger distance scales than those implied by the 
recently established year-scale variability.

Studies of the variability time scale and of coincident multiwavelength observations
are very important for discriminating between these models.

\section{Observations and analysis}
The Very Energetic Radiation Imaging Telescope Array System (VERITAS)
is a $\gamma$-ray observatory located in southern Arizona.
It is an array of four 12\,m-diameter imaging atmospheric Cherenkov telescopes,
each with a 499-pixel photomultiplier-tube camera covering a 3.5$^\circ$ field of view \citep{hol06}.
The energy range covered extends from 100\,GeV (zenith observation) up to tens of TeV.

The observations of M\,87 were carried out over 51 hours between February to April 2007
at elevations from 55$^\circ$ to 71$^\circ$.
In spring 2007, VERITAS was still in its construction phase. The M\,87 observations
were performed with an array of either three telescopes (94\% of the data) or four telescopes (6\%).
The performance of the array during this period is discussed elsewhere~\citep{mai07}.
In order to minimize systematic errors on the estimation of the flux and its variability,
we do not use the data from the fourth telescope, which was not fully calibrated at this time.
We processed all the data in the same way, using only three telescopes, even
when data from the fourth telescope were available.

Only good-quality data were considered for the analysis. Our data selection is based
on weather conditions and raw trigger rate stability.
About 90\% of the data (44\,h) pass our quality selection cuts.
The selected data have been processed with several independent analysis packages \citep{dan07}.
The results presented here come from one of these packages and
the others yield consistent results.
The analysis chain uses two-threshold image cleaning, stereo reconstruction
and cosmic-ray rejection based on standard image parameter cuts \citep{hil85}.
The energy of each shower is reconstructed using a method based on the
quantity of light measured by the telescopes and on the position of the shower-core
ground impact. The differential energy spectrum is unfolded using the correction-factor
method described in Chapter 11 of \lq\lq Statistical Data Analysis\rq\rq \citep{cow98}.

All observations were taken in \lq\lq wobble\rq\rq mode, tracking M\,87 with a 0.5$^\circ$ offset
north, south, east or west from the camera center.
In this mode the cosmic-ray background rate is estimated using
several \lq\lq mirror\rq\rq regions at the same distance from the camera center (0.5$^\circ$)
as the nominal source position.
We compare the number of events with the reconstructed shower direction in these off-source
regions (seven in this analysis) to the number of events in a region centred on the nominal
source position. The radius of each region is 0.13$^{\circ}$.
Statistical significances are calculated using the Li and Ma formula (equation 17 in \citet{li83}).

\section{VERITAS results}
We detected an excess of $259\pm44$ events from the direction of M\,87,
corresponding to a statistical significance of 5.9 standard deviations.
The analysis energy threshold for the average elevation of the M\,87 observations
(66$^{\circ}$) is approximately 250\,GeV, estimated from Monte Carlo simulations.
The time-averaged excess rate (0.10$\pm$0.02\,$\gamma$/min)
corresponds to 1.9\% of the Crab Nebula rate observed at similar elevations.

\begin{figure}
\begin{center}
\includegraphics[width=\columnwidth]{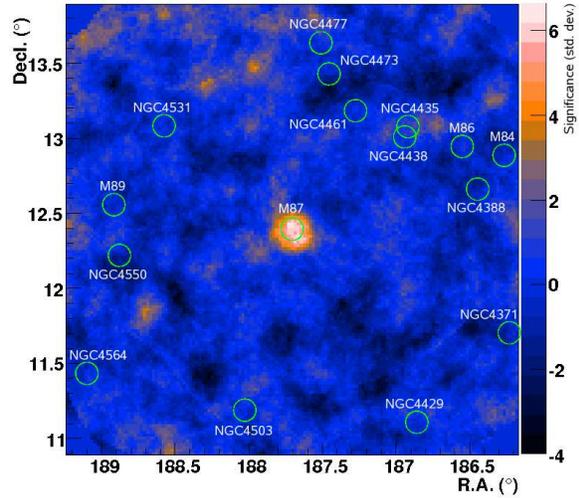}
\caption[short]{Significance map of the region surrounding M\,87 measured with VERITAS
(in units of standard deviation). Circles show the positions of the brightest galaxies of the Virgo cluster.}
\end{center}
\label{signi_map}
\end{figure}

\begin{figure}
\begin{center}
\includegraphics[width=\columnwidth]{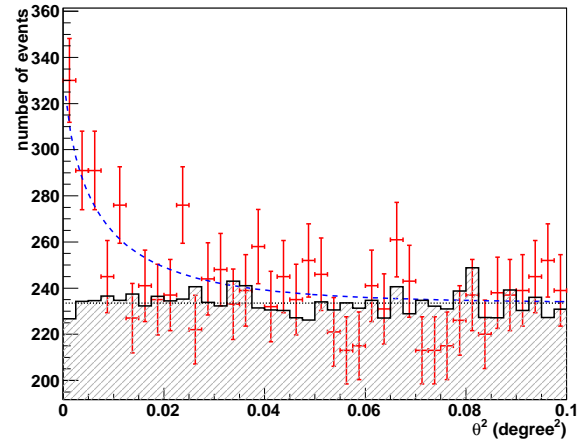}
\caption[short]{The data points with error bars show the $\theta^2$ distribution from M\,87
and the solid histogram that from off-source directions.
The dashed line is the $\theta^2$ distribution expected
for a point-like source with a background level represented by the dotted line.}
\end{center}
\label{theta2}
\end{figure}

Figure~1 shows the map of the excess significance in the
sky region of M\,87. The detected source (VER J1230+123)
is consistent with a point-like source located inside the M\,87 galaxy.
The position of the maximum of a 2D Gaussian fit of the excess is:
R.A. $12^h30^m46^s \pm4^s_{stat} \pm6^s_{sys}$,
Decl. $+12^\circ 23' 21'' \pm 50''_{stat}\pm 1'30''_{sys}$,
compatible with the position of the M\,87 core
(R.A. $12^h30^m49.4^s$, Decl. +12$^\circ 23' 28''$).
Figure~2 shows the distribution of the square of the angle $\theta$ between M\,87
and the reconstructed shower direction. The shape of the excess is compatible with
a point-like source (dashed line).
The upper limit (99\%) of the intrinsic source extension has a $4.5'$ radius.

\begin{figure}
\begin{center}
\includegraphics[width=\columnwidth]{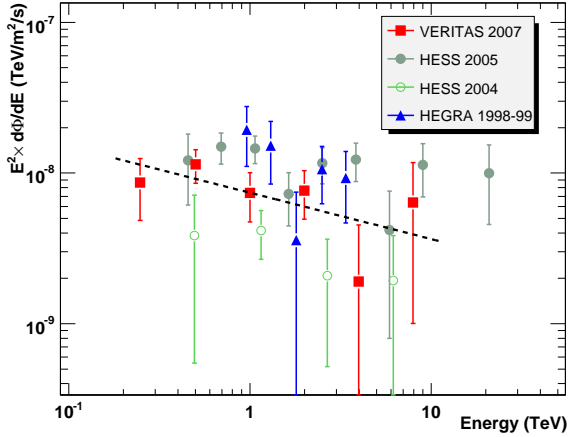}
\caption[short]{Spectral energy distribution of M\,87 measured with VERITAS in 2007
and with other experiments during the previous years.
The dashed line shows the power-law fit of the VERITAS data (see text).}
\end{center}
\label{spectrum}
\end{figure}

The differential energy spectrum has been measured from 250\,GeV to 8\,TeV.
Table~1 gives the differential flux and statistical error at different energies.
In order to keep the reconstructed energy bias below 10\%, the flux of the lowest energy
bin (250\,GeV) has been estimated using only data taken at elevation above 62$^\circ$
(34.8\,h of observation).
The M\,87 spectrum obtained is consistent with a power-law spectrum:
d$\Phi$/dE\,=$\Phi_0\cdot(E/\mbox{TeV})^\Gamma$ with a flux normalization constant
$\Phi_0=(7.4\pm1.3_{stat}\pm1.5_{sys}) 10^{-9}$m$^{-2}$s$^{-1}$TeV$^{-1}$ and
a spectral index $\Gamma=-2.31\pm0.17_{stat}\pm0.2_{sys}$.
The $\chi^2$/d.o.f. of the fit is 3.0/4.
Figure~3 shows this spectrum and its power-law fit and compares them with the previous 
spectra measured by HESS in 2004 and 2005 \citep{aha06A}.
Data points from a recent re-analysis of the HEGRA data \citep{goe06} are
also shown. The results of this HEGRA re-analysis will be used hereafter.

\begin{figure}
\begin{center}
\includegraphics[width=\columnwidth]{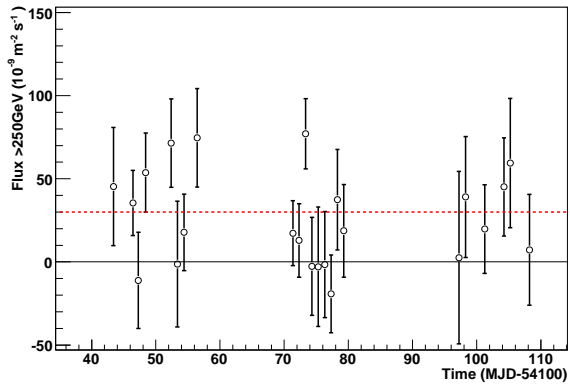}
\caption[short]{Nightly average (circles) and global average (dashed line)
of the M\,87 $\gamma$-ray flux above 250\,GeV measured with VERITAS
as a function of the day of year 2007.}
\end{center}
\label{nightlyLC}
\end{figure}

\begin{table}
\begin{center}
\caption{Differential flux of M\,87 measured with VERITAS in spring 2007}
\begin{tabular}{|c|c|c|}
\tableline\tableline
Energy & Differential flux & Significance \\
(TeV) & (m$^{-2}$s$^{-1}$TeV$^{-1}$) & (std dev.) \\
\tableline
0.25 &(1.36$\pm$0.60)\,$10^{-7}$&2.26\\
0.50 &(4.55$\pm$1.15)\,$10^{-8}$&3.97\\
1.00 &(7.39$\pm$2.65)\,$10^{-9}$&2.80\\
2.00 &(1.92$\pm$0.68)\,$10^{-9}$&2.82\\
3.98 &(1.20$\pm$1.66)\,$10^{-10}$&0.73\\
7.94 &(1.01$\pm$0.85)\,$10^{-10}$&1.19\\
\tableline
\end{tabular}
\end{center}
\end{table}

Correlation between the spectral index and the flux have been observed in BL lac
objects such as Mrk\,501 \citep{dja99} and Mrk\,421 (\citet{kre02}, \citet{alb07}).
Unfortunately, the $\gamma$-ray flux from M\,87 being so close to the sensitivity of
the present generation of atmospheric Cherenkov telescopes, the uncertainties
on the spectral index do not permit to make any statement on a possible similar behavior.

The $\gamma$-ray flux shows significant variability from one year to another.
The first panel of Figure~5 shows the integral $\gamma$-ray flux reported by the HEGRA, HESS
and VERITAS collaborations. For ease of comparison, the VERITAS flux was scaled to the
energy threshold of 730\,GeV commonly used by HEGRA and HESS.
The integral flux above 730\,GeV is estimated according
to our spectral analysis result (power-law spectrum with a spectral index $\Gamma$=-2.31):\\
$\Phi_{>\mathrm{730GeV}}$=(8.5$\pm$1.5$_{stat}\pm$1.7$_{sys}) 10^{-9}$m$^{-2}$s$^{-1}$.

The TeV $\gamma$-ray lightcurve consists of data points from different experiments.
The preliminary result for the Crab Nebula, standard candle of the TeV astronomy,
observed with VERITAS in spring 2007 \citep{cel07} is in good agreement with the flux
reported by HEGRA \citep{aha04} and HESS \citep{aha06B}. Thus, the differences observed with
different detectors comes from real variations of the M\,87 flux.

Figure~4 shows the M\,87 $\gamma$-ray flux night-by-night during the three
months of observation with VERITAS. No significant short-time-scale variability
is observed. The constant flux fit $\chi^2$ per degree of freedom (d.o.f.) is 24.3/22.
The maximum deviation from the average (MJD 54173) is only 2.2 standard deviations.

\section{X-ray / $\gamma$-ray correlation}
The All-Sky Monitor (ASM) on the Rossi X-ray Timing Explorer (RXTE)
has been monitoring the M\,87 emission between 2\,keV and 10\,keV since early 1996
\citep{lev96}. Measurements are performed with sequences of 90-second "dwells".
The light-curve data used here are the quick-look results provided by the ASM/RXTE team
on their web site\footnote{http://xte.mit.edu/asmlc/ASM.html},
dwell-by-dwell and as a daily average.

In order to check the long-term variability of the ASM/RXTE signal, we sum the data by bins
of 6\,months using the standard practice of weighting measurements by $1/\sigma^2$,
where $\sigma$ is the measurement uncertainty.
The second panel of Figure~5 shows the average ASM/RXTE rate for the first six months of each year
(i.e. the period coinciding with the observation of M\,87 with the ground-based detectors)
since 1996. The ASM/RXTE lightcurve has similar variations as the TeV lightcurve shown in the first panel.
Figure~6 shows the $\gamma$-ray flux, $\Phi_{>\mathrm{730GeV}}$, plotted versus
the ASM/RXTE rate, $X_{rate}$. Taking the error bars, which correspond to the statistical error,
into account, the linear correlation coefficient is $0.78\pm0.11$
(positive correlation confidence level $>$\,99.99\%).
The most recent analysis of the HEGRA data has been used to calculate this correlation coefficient.

A marginal correlation between the TeV $\gamma$-ray and the ASM/RXTE rate
was previously suggested by the Whipple 10\,m-telescope collaboration in 2000-01 \citep{leb03A}
and was discussed in the doctoral theses of M.~\citet{bei06} and of N.~\citet{goe06}.
Such a correlation between the TeV and hard X-ray emissions has been also detected
in the BL lac objects Mrk\,501 (\citet{dja99}, \citet{kra02}) and Mrk\,421 \citep{fos07}. The correlation
for these two objects is generally close to quadratic but depends on the energy range
considered \citep{kat05}.

We estimate the value of both the variable and the steady components of the X-ray
emission of M\,87 by fitting the TeV correlation plot with a linear relation:\\
$\Phi_{>\mathrm{730GeV}}$=(88$\pm$23)$\cdot$($X_{rate}$\,-\,(1.196$\pm$0.022))$\cdot$10$^{-9}$\\
and with a quadratic relation:\\
$\Phi_{>\mathrm{730GeV}}$=(338$\pm$180)$\cdot$($X_{rate}$\,-\,(1.142$\pm$0.042))$^2\cdot$10$^{-9}$.\\
In these two equations $\Phi_{>\mathrm{730GeV}}$ is expressed in m$^{-2}$s$^{-1}$ and $X_{rate}$ in count/s.
Both functions fit the data well with a $\chi^2$/d.o.f.$<$1.
They show that the X-ray emission consists of both a dominant steady emission (1.1-1.2\,count/s)
and a variable fraction ($<$20\%) strongly correlated with the $\gamma$-ray emission.

The constant part of the X-ray emission is dominated by the thermal emission from the $\sim$2\,keV
gaseous atmosphere of the M\,87 galactic halo \citep{for07}.
The variable part could result from synchrotron emission by high-energy particles
with cooling time less than one year. The emission site for the variable component
must have a compact size on the order of one lightyear or less ($\sim$5\,mas at the M\,87 distance),
or be moving at relativistic speed in a direction close to the line of sight, to explain the yearly
variability recorded.

The only known M\,87 regions which have such characteristics are the core and the brightest knots
of the jet. The {\sl Chandra} X-ray observatory has been used to monitor the M\,87
jet emission in the 0.2-6\,keV energy range since 2000 \citep{har06}
with enough angular resolution to measure separately the flux from the different features.
Figure~5 shows the light curves of the two dominant X-ray features which are the core and HST-1.
Since 2003, HST-1 has been flaring and it largely dominates the {\sl Chandra} X-ray flux.
However the variation of the flux from this knot does not seem to correlate well with the variation of
the ASM/RXTE signal. Between 2003 and 2005 the HST-1 X-ray emission increased by a factor of 20
whereas the variable fraction of the ASM/RXTE rate changed by less than a factor of two.
The same goes for the new flux measurement with VERITAS which clearly do not follow the X-ray lightcurve of HST-1.

The next-brightest features of the jet observed with {\sl Chandra} can explain only with difficulty
the ASM/RXTE rate variation as their fluxes are significantly weaker than the HST-1 X-ray variation.
Actually, the energy range of ASM/RXTE (2-10\,keV) is higher than that of {\sl Chandra} (0.2-6\,keV)
and the dominant features of the ASM/RXTE energy band may not be HST-1.
According to the spectral analysis of {\sl Chandra} data taken in 2000 \citep{per05},
the core was the only bright X-ray feature with a harder spectrum than HST-1.
Thus the core seems the best candidate to dominate HST-1 in hard X-rays
and provide substantial ASM/RXTE rate variation.
Moreover the confidence level for a positive linear correlation between the 6-month
average rates of ASM/RXTE and {\sl Chandra} is higher for the core (90\%) than for HST-1 (65\%).
Since the variable component of the ASM/RXTE flux is likely dominated by
X-ray emission from the core, and there is a strong correlation
between the TeV and ASM/RXTE fluxes, we conclude that the core is the most
likely candidate for the source of the TeV $\gamma$-ray emission.

It is however surprising that the large HST-1 flare in 2005 does not show in AMS/RXTE as
this implies a very soft HST-1 spectrum. There is no evidence for such a soft spectrum in
the {\sl Chandra} data. Unfortunately, during the flaring period of HST-1,
spectral analysis of the core and HST-1 seems impossible because the flux was so
strong that the bulk of the events in the image are piled, and this destroys the spectral information.
Furthermore the core is separated by only $0.86''$ from HST-1, and its {\sl Chandra} signal
can be contaminated by the HST-1 signal. As shown in \citet{har07A}, the small flare of the core
in 2005 can be almost smoothed out by subtracting 5\% of the HST-1 signal
(dotted line in the third panel of the figure~5). This is probably indicative of the uncertainties
on the core flux measurement with {\sl Chandra} during this period.

\begin{figure}
\begin{center}
\includegraphics[width=\columnwidth]{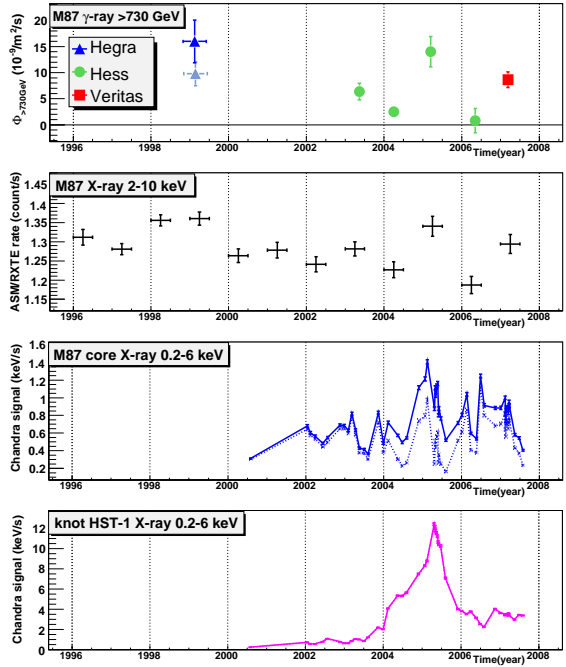}
\caption[short]{
The first panel shows the integral $\gamma$-ray flux above 730\,GeV
reported by the HEGRA, HESS and VERITAS collaborations.
Concerning the HEGRA flux, both the published result (light color) and the most recent
re-analysis result (dark color) are shown (see text for references).
The second panel shows the January-to-June average of the global X-ray emission of M\,87 
measured with ASM/RXTE in the range 2-10 keV.
The last two panels show the M87 core and the knot HST-1 X-ray emission
measured with {\sl Chandra} in the range 0.2-6 keV.
The dotted line on the third panel shows the corrected core
signal from which 5\% of the HST-1 signal were subtracted \citep{har07B}.}
\end{center}
\label{LC10Y}
\end{figure}

\begin{figure}
\begin{center}
\includegraphics[width=\columnwidth]{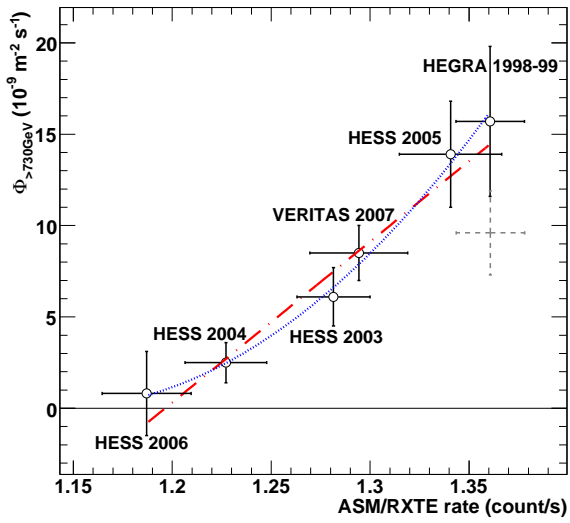}
\caption[short]{Correlation between M\,87 emission in X-ray (ASM/RXTE data: 2-10 keV)
and in $\gamma$-ray (HEGRA, HESS, VERITAS data: $>$730GeV).
The dash-dotted line is the linear correlation fit
and the dotted line shows the quadratic correlation fit.
For HEGRA, both the published flux (dashed lines) and the reanalysis result(solid lines)
are shown.
}
\end{center}
\label{year_correlation}
\end{figure}

Motivated by the long-time-scale correlation between $X_{rate}$ and $\Phi_{>730GeV}$,
we searched for a similar correlation on a 5-day scale.
Such short-time-scale correlation had previously been suggested
from earlier Whipple 10m data \citep{leb03B}.
This study was also designed to improve our sensitivity to short-time-scale variations such as
those reported by the HESS collaboration.
Again, we express the VERITAS results as the integral $\gamma$-ray flux above 730\,GeV.
Figure~7 shows $\Phi_{>730GeV}$ versus $X_{rate}$ summed on contiguous and independent 5-day-scale bins.
The confidence level for a positive correlation is only 82\%.
The statistics are too poor for us to draw any conclusion as to a possible extension of
the year-scale correlation (dash-dotted line) to the 5-day scale.
Simultaneous observations of M\,87 with VERITAS and ASM/RXTE will continue in the coming years,
providing the opportunity to investigate further such short-time-scale correlation.

\begin{figure}
\begin{center}
\includegraphics[width=\columnwidth]{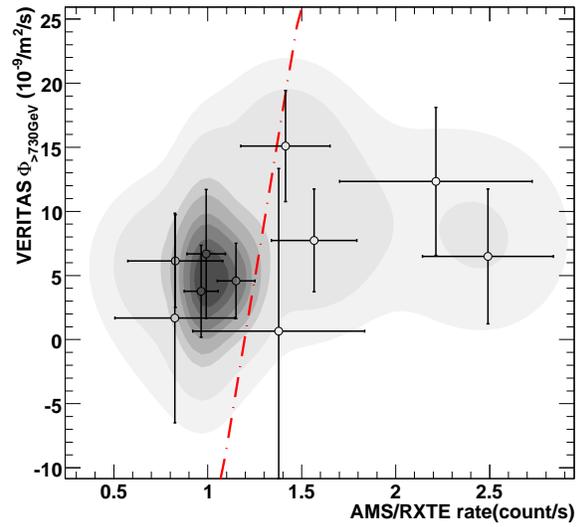}
\caption[short]{M\,87 X-ray emission (ASM/RXTE data: 2-10\,keV) plotted relative to
the very-high-energy $\gamma$-ray emission (VERITAS data: $>$730\,GeV) for contemporenous
observations. Data are binned in a 5-day intervals.
The dash-dotted line shows the linear correlation fit obtained for the year-scale correlation
in figure~6. The grey contours are the sum of 2D Gaussian functions associated with
each data point according to its uncertainty.
}
\end{center}
\label{5day_correlation}
\end{figure}

\section{Conclusion}
The VERITAS collaboration observed M\,87 in spring 2007 with an
array of three telescopes. The analysis of the 44 hours of data provides a 5.9
standard deviation detection of $\gamma$-ray emission above 250\,GeV. The energy
spectrum is fitted well by a power-law function:
d$\Phi$/dE\,=$(7.4\pm1.3_{stat}\pm1.5_{sys})\times$ (E/TeV)$^{(-2.31\pm0.17_{stat}\pm0.2_{sys})} 10^{-9}$m$^{-2}$s$^{-1}$TeV$^{-1}$.
The comparison of our result with the spectra reported by the HESS and the HEGRA collaborations
does not indicate a variability of the spectral index.

In most TeV $\gamma$-ray emission models, a correlation between the $\gamma$-ray and X-ray
flux is expected and the relationship can be used to distinguish between models.
We have found strong evidence for such a correlation from M\,87.
We have shown a year-scale correlation between the $\gamma$-ray flux above 730\,GeV, recorded by
imaging atmospheric Cherenkov telescopes during the last ten years, and the X-ray flux in the energy range 2-10\,keV recorded with ASM/RXTE. Both linear and quadratic functions fit this correlation well.
The poor correlation between the ASM/RXTE rate variations and the dominant X-ray features (HST-1) observed
by {\sl Chandra} in the 0.2-6\,keV energy band is surprising but, in the light of the correlation with the
TeV flux, it suggests HST-1 is unlikely to be the main source of the $\gamma$-ray emission.
If the core indeed has a harder spectrum than HST-1, it could dominate in the
ASM/RXTE energy range, and be the most likely site of the detected TeV emission.
HESS observations of M\,87 in 2005 were suggestive of a 2-day-scale variability.
The VERITAS observations in 2007 showed it to be in a lower state and, even
with a short-time-scale X-ray correlation study, the statistics of this observation
are insufficient to confirm such rapid variability.
Further observations and multiwavelength campaigns on M\,87 at a range of activity states
will prove important in tracking down the site of the high-energy emission, determining
the particle acceleration mechanism and the relation to the blazar class of high-energy objects.

\acknowledgments

The authors are grateful to the ASM/RXTE teams at MIT and at the RXTE SOF and GOF 
at NASA's GSFC and specially to Ron Remillard who provided the ASM/RXTE data.
We also thank Dan Harris for his comments on the {\sl Chandra} observations.

VERITAS is supported by grants from the U.S. Department of Energy, the
U.S. National Science Foundation and the Smithsonian Institution, by
NSERC in Canada, by PPARC in the U.K. and by Science Foundation Ireland.



{\it Facilities:} \facility{VERITAS}, \facility{HEGRA}, \facility{HESS},
 \facility{RXTE (ASM)}, \facility{CXO (ACIS)}.





\begin{thebibliography}{}
\bibitem[Aharonian et al.(2003)]{aha03} 
Aharonian F. et al. (HEGRA coll.) 2003, A\&A , 403, L1
\bibitem[Aharonian et al.(2004)]{aha04} 
Aharonian F. et al. (HEGRA coll.) 2004,\apj, 614, 897
\bibitem[Aharonian et al.(2006A)]{aha06A} 
Aharonian F. et al. (HESS coll.) 2006A, Science , 314, 1424
\bibitem[Aharonian et al.(2006B)]{aha06B} 
Aharonian F. et al. (HESS coll.) 2006B, A\&A , 457, 899
\bibitem[Albert et al.(2007)]{alb07} 
Albert et al (MAGIC coll.) 2007, \apj, 663 125
\bibitem[Baltz et al.(2000)]{bal00}
Baltz, E. A. et al. 2000, Phys. Rev. D61, 023514
\bibitem[Beilicke (2006)]{bei06}
Beilicke, M. 2006, thesis on Dissertation.de ISBN 3-86624-112-7 
\bibitem[Bicknell \& Begelman(1996)]{bic96}
Bicknell, G. V. \& Begelman, M. C. 1996, \apj, 467, 597
\bibitem[Biermann et al.(2001)]{bie01} 
Biermann. et al. 2001, \textit{Physics and astrophysics of UltraHigh-Energy Cosmic Rays},
M. Lemoine \& G. Sigl (Berlin: springer).
\bibitem[Biretta et al.(1999)]{bir99}
Biretta, J.A., Sparks, W. B. \& Macchetto, F. 1999, \apj, 520, 621
\bibitem[Celik et al.(2007)]{cel07} 
Celik, O. et al. (VERITAS coll.) 2007, in Proc. 30th ICRC, M\'erida, Mexico, 1209
\bibitem[Cheung et al.(2007)]{che07}
Cheung, C. C., Harris, D. E. \& Stawarz, L. 2007, \apj, 663, L65
\bibitem[Cowan(1998)]{cow98}
Cowan, G. 1998, \textit{\lq\lq Statistical Data Analysis\rq\rq}, Oxford University Press
\bibitem[Curtis(1918)]{cur18}
Curtis, H. D. 1918, Publ. Lick Obs., 13, 9
\bibitem[Daniel et al.(2007)]{dan07} 
Daniel, M. K. et al. (VERITAS coll.) 2007, in Proc. 30th ICRC, M\'erida, Mexico, 283
\bibitem[Djannati et al.(1999)]{dja99} 
Djannati-Ata\"i A. et al. (CAT coll.) 1999, A\&A, 350, 17 
\bibitem[Fanaroff \& Riley(1974)]{fan74} 
Fanaroff, B. L. \& Riley, J. M.  1974, \mnras, 167, 31P
\bibitem[Forman et al.(2007)]{for07} 
Forman, W. et al. 2007, \apj, 665, 1057
\bibitem[Fossati et al.(2007)]{fos07} 
Fossati, G. et al. 2007, accepted by \apj, arXiv:0710.4138
\bibitem[Georganopoulos et al.(2005)]{geo05}
Georganopoulos, M., Perlman, E. S. \& Kazanas D. 2005, \apj, 634, L33
\bibitem[Ghisellini et al.(2005)]{ghi05}
Ghisellini, G., Tavecchio, F. \& Chiaberge, M. 2005, A\&A, 432, 401
\bibitem[G\"otting (2006)]{goe06}
G\"otting, N. 2006, thesis on Dissertation.de ISBN 3-86624-243-2
\bibitem[Harris et al.(2006)]{har06}
Harris, D. E. et al. 2006,  \apj, 640, 211
\bibitem[Harris et al.(2007)]{har07A}
Harris, D. E. et al. 2007, arXiv:0707.3124
\bibitem[Harris (2007)]{har07B}
Harris, D. E. 2007, private communication.
\bibitem[Hillas (1985)]{hil85}
Hillas, M. 1985, in Proc. 19th ICRC, La Jolla, USA, 3, 445
\bibitem[Holder et al.(2006)]{hol06}
Holder, J. et al. (VERITAS coll.) 2006, Astropart. Phys. 25, 391
\bibitem[Honda \& Honda (2007)]{hon07}
Honda, M. \& Honda, Y. S. 2007,  \apj, 654, 885
\bibitem[Katarzynski et al.(2005)]{kat05}
Katarzynski, K. et al. 2005, A\&A 433, 479
\bibitem[Krawczynski et al.(2002)]{kra02}
Krawczynski, H., Coppi, P. S. 2002, \& Aharonian, F. A. \mnras, 337, 721
\bibitem[Krennrich et al.(2002)]{kre02}
Krennrich, F. et al. (Whipple 10m coll.) 2002, \apj, 575, L9
\bibitem[LeBohec et al.(2003A)]{leb03A}
Le Bohec, S. et al. (Whipple 10m coll.) 2003A, \apj, 610, 156
\bibitem[LeBohec et al.(2003B)]{leb03B}
Le Bohec, S. et al. (Whipple 10m coll.) 2003B, in Proc. 28th ICRC, Tsukuba, Japan
\bibitem[Lenain et al.(2007)]{len07}
Lenain, J.-P. et al. 2007, accepted by A\&A, arXiv:0710.2847
\bibitem[Levine et al.(1996)]{lev96}
Levine, M. et al. 1996, \apj, 469, L33
\bibitem[Li \& Ma(1983)]{li83}
Li, T. P. \& Ma, Y. Q. 1983, \apj, 272, 317
\bibitem[Perlman \& Wilson (2005)]{per05}
Perlman, E. S. \& Wilson, A. S. 2005, \apj, 627, 140
\bibitem[Macchetto et al.(1997)]{mac97}
Macchetto, F. et al. 1997, \apj, 489, 579
\bibitem[Maier et al.(2007)]{mai07} 
Maier, G. et al. (VERITAS coll.) 2007, in Proc. 30th ICRC, M\'erida, Mexico, 810
\bibitem[Neronov \& Aharonian(2007)]{ner07} 
Neronov, A. \& Aharonian, F. A. 2007, \apj, 671, 85
\bibitem[Reimer et al.(2004)]{rei04}
Reimer, A., Protheroe, R. J. \& Donea, A.-C. 2004, A\&A, 419, 89
\bibitem[Rieger \& Aharonian(2008)]{rie08}
Rieger, F. M. \& Aharonian, F. A. 2008, accepted by A\&A, arXiv:0712.2902
\bibitem[Stawarz et al.(2003)]{sta03}
Stawarz, L., Sikora, M. \& Ostrowski, M. 2003, \apj, 597, 186
\bibitem[Stawarz et al.(2006)]{sta06}
Stawarz, L. et al. 2006, \mnras, 370, 981
\bibitem[Tavecchio \& Ghisellini(2008)]{tav08}
Tavecchio, F. \& Ghisellini, G. 2008, accepted by \mnras, arXiv:0801.0593
\bibitem[Urry \& Padovani(1995)]{urr95}
Urry, C. M. \& Padovani, P., \pasp, 107, 803

\end{thebibliography}
\end{document}